\documentclass[]{spie}  %>>> use for US letter paper
\usepackage{amsmath,amsfonts,amssymb}
\usepackage{graphicx}
\usepackage{color}
\usepackage{gensymb}
\usepackage{booktabs}
\usepackage{overpic}

\newcommand{\figref}[1]{Fig.~\ref{#1}}
\newcommand{\tabref}[1]{Tab.~\ref{#1}}
\newcommand{\secref}[1]{$\S$~\ref{#1}}

\def\spider{{\sc spider}}
\def\planck{{\sc planck}}
\def\biceptwo{{\sc bicep-2}}
\def\keck{{\sc keck}}
\def\cmb{{\sc cmb}}
\def\spiderone{{\sc spider-1}}
\def\spidertwo{{\sc spider-2}}
\def\advact{{\sc Advanced ACTPol}}
\def\micron{$\mu$m}
\def\psat{$P_{\mathrm{sat}}$}
\def\tc{$T_{\mathrm{c}}$}
\def\rn{$R_{\mathrm{n}}$}
\def\tb{$T_{\mathrm{b}}$}

\def\apj{Astrophysical Journal}

\def\procspie{Proceedings of SPIE}
\def\jcap{Journal of Cosmology and Astroparticle Physics}
\title{Design of 280~GHz feedhorn-coupled TES arrays for the balloon-borne polarimeter SPIDER}

\author[a]{Johannes~Hubmayr}
\author[a]{Jason~E.~Austermann}
\author[a]{James~A.~Beall}
\author[a]{Daniel~T.~Becker}
\author[b]{Steven~J.~Benton}
\author[b]{A.~Stevie~Bergman}
\author[c]{J.~Richard~Bond}
\author[d]{Sean~Bryan}
\author[a]{Shannon~M.~Duff}
\author[e]{Adri~J.~Duivenvoorden}
\author[f]{H.~K.~Eriksen}
\author[g]{Jeffrey~P.~Filippini}
\author[b]{Aurelien~A.~Fraisse}
\author[h]{Mathew~Galloway}
\author[b]{Anne~E.~Gambrel}
\author[i]{K.~Ganga}
\author[a]{Arpi~L.~Grigorian}
\author[g]{Riccardo~Gualtieri}
\author[e]{Jon~E.~Gudmundsson}
\author[h]{John~W.~Hartley}
\author[j]{M.~Halpern}
\author[a]{Gene~C.~Hilton}
\author[b]{William~C.~Jones}
\author[k]{Jeffrey~J.~McMahon}
\author[l]{Lorenzo~Moncelsi}
\author[m]{Johanna~M.~Nagy}
\author[h]{C.~B.~Netterfield}
\author[g]{Benjamin~Osherson}
\author[h]{Ivan~Padilla}
\author[b]{Alexandra~S.~Rahlin}
\author[f]{B.~Racine}
\author[m]{John~Ruhl}
\author[f]{T.~M.~Ruud}
\author[m]{J.~A.~Shariff}
\author[n]{J.~D.~Soler}
\author[b]{Xue~Song}
\author[a]{Joel~N.~Ullom}
\author[a]{Jeff~Van~Lanen}
\author[a]{Michael~R.~Vissers}
\author[f]{I.~K.~Wehus}
\author[m]{Shyang~Wen}
\author[j]{D.~V.~Wiebe}
\author[b]{Edward~Young}

\authorinfo{Contribution of NIST not subject to copyright.}
\affil[a]{NIST, Boulder, CO USA}
\affil[b]{Princeton University, Jadwin Hall, Princeton, NJ USA}
\affil[c]{CITA, University of Toronto, Toronto, ON, Canada}
\affil[d]{Arizona State University, Tempe, AZ}
\affil[e]{The Oskar Klein Centre for Cosmoparticle Physics, Stockholm University, Stockholm, Sweden}
\affil[f]{Institute of Theoretical Astrophysics, University of Oslo, Oslo, Norway}
\affil[g]{University of Illinois at Urbana-Champaign Urbana, IL USA}
\affil[h]{University of Toronto, Toronto, ON, Canada}
\affil[i]{APC, University of Paris-Diderot, Obs. de Paris, Sorbonne Paris Cité, France}
\affil[j]{University of British Columbia, Vancouver, BC, Canada}
\affil[k]{University of Michigan, Ann Arbor, MI USA}
\affil[l]{California Institute of Technology, Pasadena, CA USA}
\affil[m]{Case Western Reserve University, Cleveland, OH USA}
\affil[n]{Laboratoire AIM, Paris-Saclay, Université Paris Diderot, Cedex, France}

\begin{document} 
\maketitle

\begin{abstract}
We describe 280~GHz bolometric detector arrays that instrument the balloon-borne polarimeter \spider. 
A primary science goal of \spider\ is to measure the large-scale B-mode polarization of the cosmic microwave background (\cmb) in search of the cosmic-inflation, gravitational-wave signature.   
280~GHz channels aid this science goal by constraining the level of B-mode contamination from galactic dust emission.    
We present the focal plane unit design, which consists of a 16$\times$16 array of conical, corrugated feedhorns coupled to a monolithic detector array fabricated on a 150~mm diameter silicon wafer.
Detector arrays are capable of polarimetric sensing via waveguide probe-coupling to a multiplexed array of transition-edge-sensor (TES) bolometers.
The \spider\ receiver has three focal plane units at 280~GHz, which in total contains 765 spatial pixels and 1,530 polarization sensitive bolometers. 
By fabrication and measurement of single feedhorns, we demonstrate 14.7$^{\circ}$ FHWM Gaussian-shaped beams with $<$1\% ellipticity in a 30\% fractional bandwidth centered at 280 GHz. 
We present electromagnetic simulations of the detection circuit, which show 94\%  band-averaged, single-polarization coupling efficiency, 3\% reflection and 3\% radiative loss.
Lastly, we demonstrate a low thermal conductance bolometer, which is well-described by a simple TES model and exhibits an electrical 
noise equivalent power (NEP)~=~2.6~$\times$~10$^{-17}$ W/$\sqrt{\mathrm{Hz}}$, consistent with the phonon noise prediction.
\end{abstract}


%%%%%%%%%%%%%%%%%%%%%%%%%%%%%%%%%%%%%%%%%%%%%%%%%%%%
%%%%%%%%%%%%%%%%%%%%%%%%%%%%%%%%%%%%%%%%%%%%%%%%%%%%
%%%%%%%%%%%%%%%%%%%%%%%%%%%%%%%%%%%%%%%%%%%%%%%%%%%%
\section{Introduction}
\label{sec:intro}  

% paragraph on general BG of cmb polarization: interest, relation to inflation.  Describe measurement details including foregrounds
Measurements of the polarization of the cosmic microwave background (\cmb) currently receive significant attention, as they are a known probe of cosmic inflation.
Several experiments target the measurement of the signatures of inflation 
\cite{fraisse2013spider, reichborn2010ebex, chuss2010piper, henderson2016advanced, benson2014spt3g, ghribi2014latest, staniszewski2012keck, suzuki2016polarbear, essinger2014class, aiola2012lspe}. 
Currently only upper limits exist on the tensor-to-scalar ratio $r$ \cite{bk62016improved}, a parameter that constrains inflationary physics.
From \planck\ and \biceptwo/\keck\ measurements \cite{bicep22014detection,bkp2015joint}, a new appreciation of polarized foregrounds has been gained.
A robust detection of inflation signatures will require foreground characterization, which in turn requires broad spectral coverage.
Earth's atmosphere limits the available observation frequency bands, and so balloon observations are a logical platform choice for increased spectral coverage.   

% paragraph on spider
\spider\ is a balloon-borne instrument that aims to measure or constrain $r~\sim~0.03$.
The first flight of \spider, which we refer to as \spiderone, occurred in January 2015.  
In this flight, \spider\ mapped $\sim$~7\% of the sky at 95~GHz and 150~GHz.    
The upcoming flight of \spider, which we refer to as \spidertwo, includes three 280~GHz focal planes and three of the previously flown arrays.  
In total, \spidertwo\ has 765 dual-polarization-sensitive spatial pixels and 1,530 transition edge sensor (TES) bolometers at 280~GHz. 
These new arrays are used to characterize polarized dust foregrounds.
We utilize the silicon feedhorn-coupled TES polarimeter architecture \cite{yoon2009feedhorn}, which was initially developed for the 150~GHz band.  
Arrays of this architecture at 150~GHz and multichroic arrays at 90/150~GHz have been deployed in ground-based systems \cite{austermann2012,thornton2016actpol,datta2016design,essinger2010abs}.    
\spidertwo\ is the first implementation of the technology at 280~GHz and the first implementation optimized for balloon-based observing. 

% paragraph on the emphasis of this paper.
This paper presents the design of the \spidertwo, 280~GHz focal plane arrays.
\secref{sec:fp} overviews the focal plane unit.  
\secref{sec:feeds} presents the feedhorn array design as well as measured performance.  
\secref{sec:det} describes the polarization sensitive bolometric array, and \secref{sec:det_design} details the pixel design with particular emphasis on the electromagnetic simulations of circuit elements.    
\secref{sec:bolo_study} discusses measurements of a prototype bolometer, suitable for balloon-based observing that is well-described by a simple TES bolometer model.

%%%%%%%%%%%%%%%%%%%%%%%%%%%%%%%%%%%%%%%%%%%%%%%%%%%%
%%%%%%%%%%%%%%%%%%%%%%%%%%%%%%%%%%%%%%%%%%%%%%%%%%%%
%%%%%%%%%%%%%%%%%%%%%%%%%%%%%%%%%%%%%%%%%%%%%%%%%%%%
\section{Focal Plane Overview}
\label{sec:fp}

% Fig1: spider-FPU
\begin{figure*}[t]
\begin{overpic}[width=6.5in]{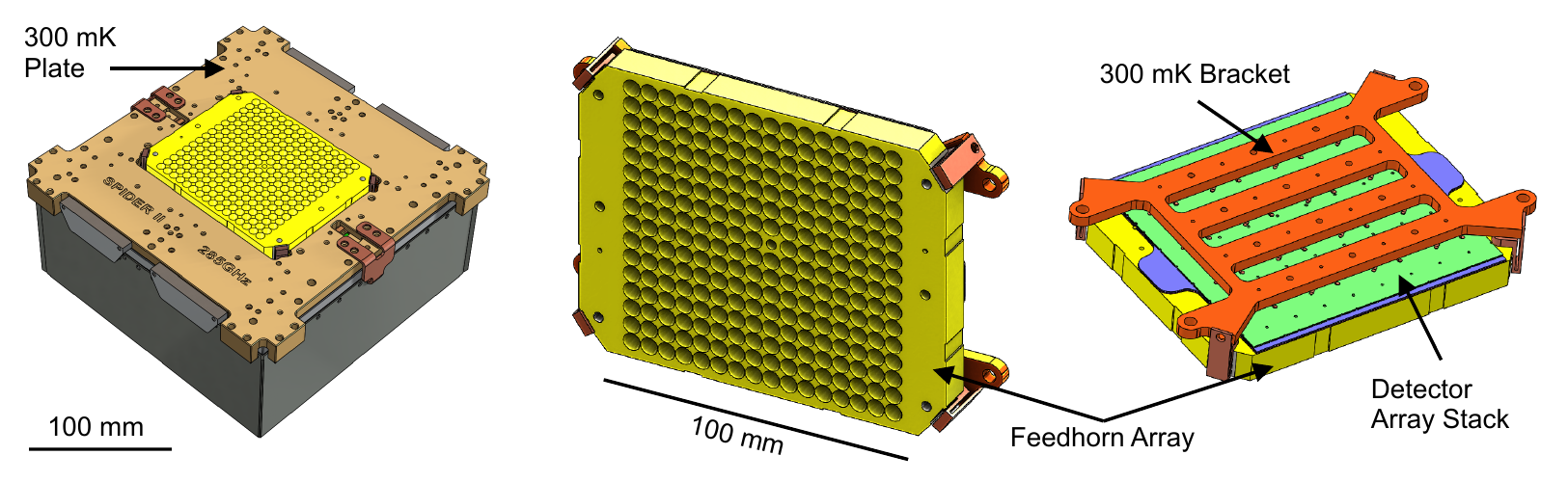}
\put(14,31){\normalsize \textcolor{black}{\bf (a)}}
\put(47,31){\normalsize \textcolor{black}{\bf (b)}}
\put(80,31){\normalsize \textcolor{black}{\bf (c)}}
\end{overpic}
\caption{\spidertwo\ 280~GHz focal plane design. {\bf (a)} Focal plane unit.  
The feedhorn array is visible in the center and is surrounded by a 300~mK copper plate.  
SQUID multiplexing hardware folds behind the array by use of flexible cabling, and is located within the grey box beneath the 300~mK plate.
{\bf (b)} Front view of sensor array assembly.
{\bf (c)} Rear view of sensor array assembly.  
The detector array stack presses against the feedhorn array by use of BeCu springs located between the 300~mK bracket and the detector array stack.}
\label{fig:focalplane}
\end{figure*}

% paragraph placing focal plane in context with the rest of the instrument.
The \spider\ instrument contains six cryogenic telescopes housed in a single receiver and has previously been described in a number of publications \cite{montroy2006spider,crill2008spider,filippini2010spider,rahlin2014preflight,gudmundsson2015thermal}.
Each telescope consists of two 4~K cooled lenses and a bolometric array at the focal plane cooled to 300~mK, which is read out with a time division SQUID multiplexer \cite{deKorte2003time}.
Three of the six arrays in \spidertwo\ are designed for the 280~GHz band, and the description of these arrays is the focus of this paper.  
The design of the 280~GHz telescopes for \spidertwo\ has been driven by the desire to conform to the existing \spiderone\ instrumentation and footprint within the receiver.
As such, the new arrays have been optimized to couple to the existing 512 channels of readout, which dictates a maximum of 256 dual-polarization-sensitive spatial pixels.

% paragraph on focal plane
\figref{fig:focalplane} shows the design of the focal plane unit, which implements NIST-developed, feedhorn-coupled TES polarimeter arrays \cite{yoon2009feedhorn}.  
In this architecture, a monolithic feedhorn array made of Au-plated silicon couples to a detector array that realizes dual-polarization-sensitivity via planar orthomode transducers (OMTs) 
connected to multiplexed transition-edge-sensor (TES) bolometers.    
A \spidertwo\ 280~GHz array consists of a 5.778~mm pitch, 16$\times$16 square array of conical, corrugated feedhorns (described in \secref{sec:feeds}), which mate to a 3-part silicon detector array stack (described in \secref{sec:det}).
Passbands are defined by waveguide cutoff, integrated on the end of the feedhorns, and a 315~GHz low pass filter \cite{ade2006review} mounted in front of the array.
There are 255 optically active pixels, as one of the central horn locations is used in feedhorn construction.
The detectors behind this pixel are the dark detectors on the array, used for systematic checks.  
The detector and feedhorn assembly are mechanically held by a BeCu spring-clamp mechanism, a technique which has been used in previous implementations of the technology \cite{henning2012feedhorn,grace2014actpol}.
Detector wiring exits the top and bottom of the wafer and couples to the SQUID readout through superconducting aluminum flexible cables, which enable the readout modules to fold behind the focal plane as in \spiderone\ \cite{runyan2010design}.  
The feedhorns, detector arrays, and SQUID multiplexers are fabricated in the NIST Boulder microfabrication facility.

%%%%%%%%%%%%%%%%%%%%%%%%%%%%%%%%%%%%%%%%%%%%%%%%%%%%
%%%%%%%%%%%%%%%%%%%%%%%%%%%%%%%%%%%%%%%%%%%%%%%%%%%%
%%%%%%%%%%%%%%%%%%%%%%%%%%%%%%%%%%%%%%%%%%%%%%%%%%%%
\section{Feedhorn Array}
\label{sec:feeds}

% Fig2: feedhorns 
\begin{figure*}[t]
\begin{overpic}[width=6.5in]{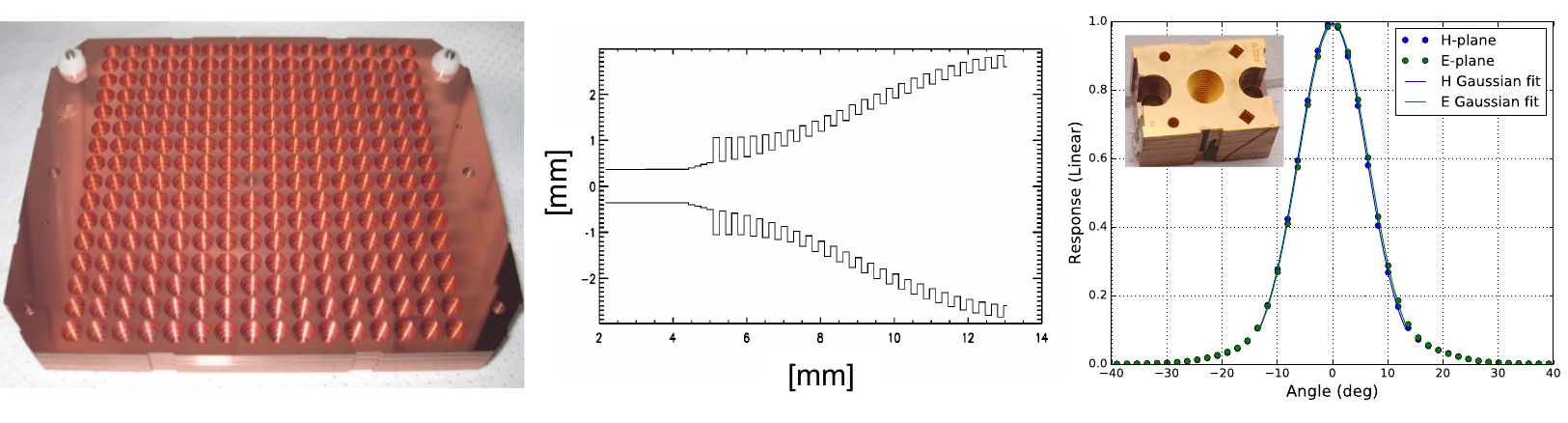}
\put(16,27){\normalsize \textcolor{black}{\bf (a)}}
\put(51,27){\normalsize \textcolor{black}{\bf (b)}}
\put(83.5,27){\normalsize \textcolor{black}{\bf (c)}}
\end{overpic}
\caption{Feedhorn array.
{\bf (a)} Photograph of assembled feedhorn array with Ti/Cu seed layer.  
{\bf (b)} Corrugation profile.
{\bf (c)} Measurement and Gaussian fit to the E-plane and H-plane angular response of a single pixel feed, shown in the inset.  
Points are the average measurement over the frequency band, and solid lines are the Gaussian fit.}
\label{fig:feeds}
\end{figure*}

\figref{fig:feeds}a shows an assembled 255-pixel array of 280~GHz corrugated feedhorns.  
The corrugation profile is created by stacking 333~\micron\ thick silicon plates that use a two-step etch process to create 167~\micron\ corrugation slots and teeth.
Fabrication details of the silicon platelets are described in previous publications \cite{britton2010corrugated, nibarger2012}.
Each platelet receives a Ti/Cu seed layer.
The platelets are stacked and aligned to 10~\micron\ plate-to-plate precision by use of dowel pins and pin guides.
%We verify wafer-to-wafer alignment by photographing edge features optical edge alignment (\comment{how do I really describe this?)}
We subsequently Au electroplate the stack to form a continuous metal surface.  

In \figref{fig:feeds}b we present the corrugation profile.  
The 5.195~mm diameter output aperture is sized to achieve $-10$~dB edge taper coupling to f/2.2 optics.  
The corrugation profile geometry follows the design rules described in Granet and James \cite{granet2005design}, and is further optimized using both mode-matching software and full electromagnetic simulations. 
We optimize over the single-mode bandwidth 242--315~GHz.  % Do I need a comment on the length of the feedhorn?  
The low edge of the band is defined by a 725~\micron\ diameter circular waveguide, integrated into the detector end of the feedhorn.  

\figref{fig:feeds}c shows the E-plane and H-plane angular response of a single-pixel feed (photograph inset) that has been fabricated and characterized to validate the design.  
The data presented are the band-averaged response of a series of microwave vector network analyzer measurements over the frequency range 245--335~GHz, taken in 10~GHz steps.
In these measurements, a receiving horn sweeps about the phase center of the device under test.  
Further details of the measurement system can be found in a previous publication \cite{britton2010corrugated}.
E- and H-plane measurements are well fit by Gaussian profiles with best-fit Gaussian response showing 1\% residuals over the  0~dB to $-10$~dB vertical range.  
The beam waist matches the designed 14.7$^{\circ}$ FHWM, and the E-plane and H-plane beam waists differ by 0.6\%.  

%%%%%%%%%%%%%%%%%%%%%%%%%%%%%%%%%%%%%%%%%%%%%%%%%%%%
%%%%%%%%%%%%%%%%%%%%%%%%%%%%%%%%%%%%%%%%%%%%%%%%%%%%
%%%%%%%%%%%%%%%%%%%%%%%%%%%%%%%%%%%%%%%%%%%%%%%%%%%%
\section{Detector Array}
\label{sec:det}
% Fig3: Detector array
\begin{figure*}[t]
\vspace{0.25in}
\begin{overpic}[width=6.5in]{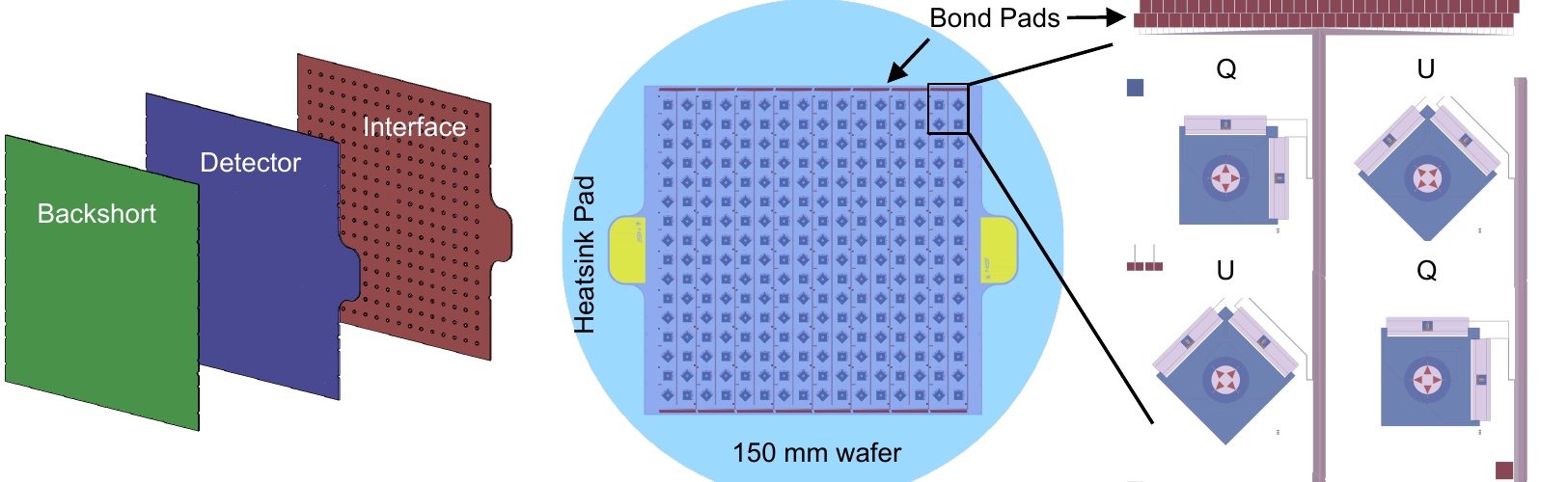}
\put(16,33){\normalsize \textcolor{black}{\bf (a)}}
\put(50,33){\normalsize \textcolor{black}{\bf (b)}}
\put(83.5,33){\normalsize \textcolor{black}{\bf (c)}}
\end{overpic}
\caption{Detector array stack.
{\bf (a)} The detector array stack comprises three silicon parts: the backshort, detector array, and feedhorn interface wafer.    
{\bf (b)} Detector array design drawn on a 150~mm diameter wafer.
{\bf (c)} Zoom-in on four pixels showing Q/U Stokes parameter pixel neighbors, wiring bundles, and bond pads.  Details of the pixel architecture are presented in \figref{fig:det_design}.}
\label{fig:det_stack}
\end{figure*}

% Figure4: detector montage
\begin{figure}[b]
   \begin{center}
   \begin{tabular}{c} %% tabular useful for creating an array of images 
   \includegraphics[height=3in]{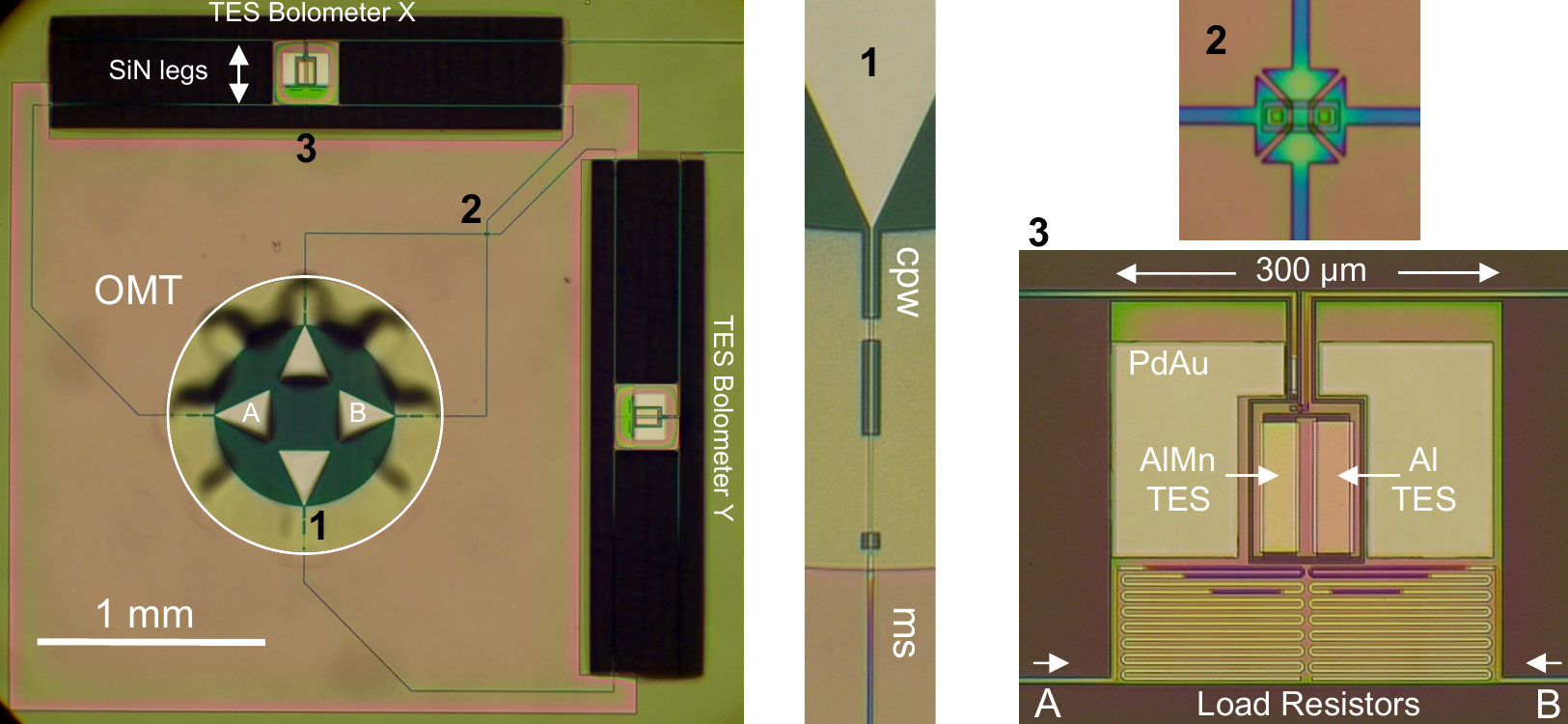}
   \end{tabular}
   \end{center}
   \caption[example] 
%>>>> use \label inside caption to get Fig. number with \ref{}
  { \label{fig:det_design} 
Pixel architecture.  Optical micrograph of a \spidertwo\ detector pixel and zoom-in on elements {\bf 1} co-planar waveguide (CPW) to microstrip (MS) transition,
{\bf 2} microwave cross-under, and {\bf 3} transition-edge-sensor (TES) bolometer.  The bolometer contains two sensors wired in series: a 420~mK AlMn sensor for science observations and 
a $\sim$~1.6~K Al sensor to extend the bolometer dynamic range for lab-based measurements.}
   \end{figure}

% paragraph on array and integration
As illustrated in \figref{fig:det_stack}a, the detector assembly consists of three silicon parts: a waveguide interface, the detector wafer, and a backshort.  
The stack realizes the architecture of polarization sensitive probes inserted into circular waveguides  with a reflective backshort.
The three parts are aligned using a custom jig with dowel pin holes and guide pins.  
Dimensional tolerance on the dowel pin diameter, circular etches in the silicon parts, and Au electroplating thickness dictates 16~\micron\ maximum misalignment.  
The aligned stack is subsequently glued around its perimeter.  

% paragraph on detector array
\figref{fig:det_stack}b shows the detector wafer layout on top of a standard 150~mm diameter silicon wafer.  
We checker-board tile the 16$\times$16 dual-polarization-sensitive pixel array with 0$^{\circ}$ and 45$^{\circ}$ oriented polarimeter axes, such that neighboring pixels in an azimuth scan measure Stokes parameters I, Q, and U.  
The 2$\times$2 pixel zoom-in of \figref{fig:det_stack}c explicitly shows this implementation.
Each detector column couples to a 1$\times$32 time-division SQUID multiplexer chip.
The entire array requires 16 columns of readout.
Heatsink pads exist on either side of the array and are also used as the mounting points for neutron transmutation-doped (NTD) thermometers used to monitor the detector wafer temperature.

For fabrication details we refer the reader to Duff et al. \cite{duff2016advanced}, which details the near identical fabrication process of the \advact\ detector arrays. 
The notable differences in this implementation are the use of 420~mK AlMn sensors; the implementation of a series, dynamic range extending Al sensor; and the use of 1~\micron\ low stress nitride for the OMT membrane probes 
and bolometer thermal isolation.  
%The former is required to couple to the \spider\ 300~mK cooler, and the latter is to achieve the lower bolometer saturation power target, 
%the purpose of which is discussed in \secref{sec:bolo_study}.

%%%%%%%%%%%%%%%%%%%%%%%%%%%%%%%%%%%%%%%%%%%%%%%%%%%%
%%%%%%%%%%%%%%%%%%%%%%%%%%%%%%%%%%%%%%%%%%%%%%%%%%%%   
\section{Detector Design}
\label{sec:det_design}

Each detector comprises an integrated superconducting circuit with elements for polarization diplexing and power sensing.  
The circuit elements of a pixel are a planar ortho-mode transducer (OMT); 
a co-planar waveguide (CPW) to microstrip transition; 
a microwave cross-under; 
and two transition edge sensor bolometers, which contain niobium to gold microstrip transitions to deposit power in the bolometers.  
\figref{fig:det_design} shows an optical micrograph of a fabricated single pixel and highlights the individual circuit elements. 

%%%%%%%%%%%%%%%%%%%%%%%%%%%%%%%%%%%%%%%%%%%%%%%%%%%%
\subsection{Electromagnetic design and performance}
\subsubsection{Ortho-mode Transducer (OMT)}
\label{sec:omt}

% Fig5: OMT model and results 
\begin{figure*}[t]
\begin{overpic}[width=6.5in]{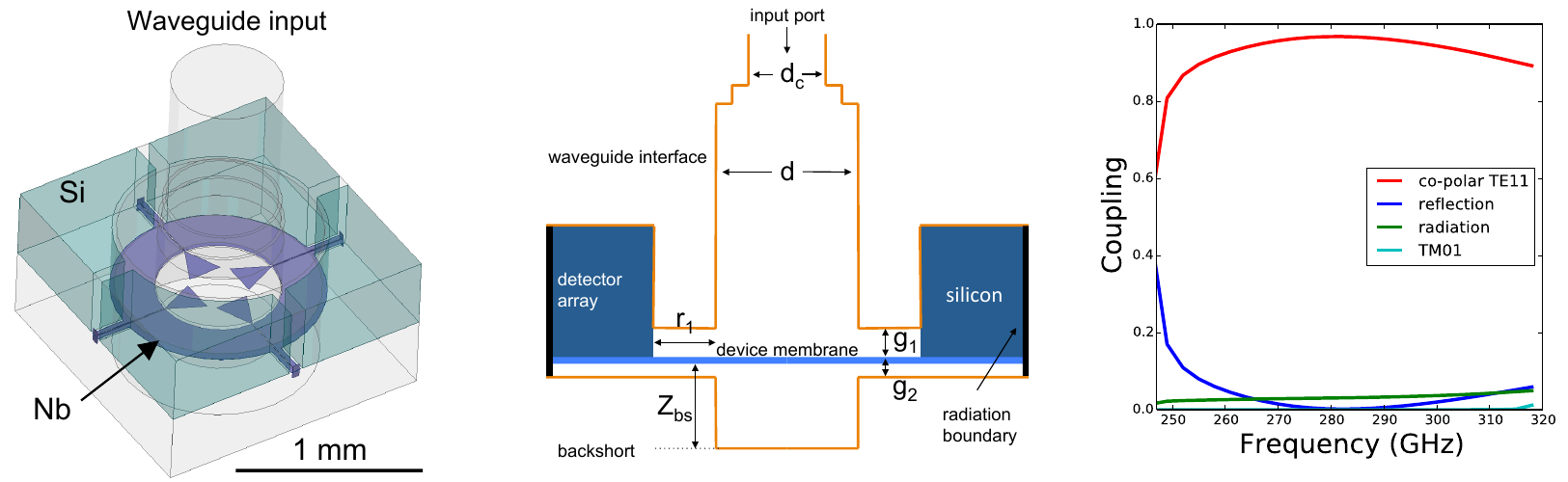}
\put(0,27){\normalsize \textcolor{black}{\bf (a)}}
\put(33,27){\normalsize \textcolor{black}{\bf (b)}}
\put(67,27){\normalsize \textcolor{black}{\bf (c)}}
\end{overpic}
\caption{280~GHz ortho-mode transducer (OMT) electromagnetic modeling.
{\bf (a)} HFSS simulation volume.
{\bf (b)} Cross-section of model (not to scale).
{\bf (c)} OMT-coupling simulations results over the band 242~GHz -- 315~GHz}
\label{fig:omt_model}
\end{figure*}
 
% Requires the booktabs if the memoir class is not being used
\begin{table}[b]
   \centering
   %\topcaption{Table captions are better up top} % requires the topcapt package
   \begin{tabular}{@{} lcr @{}} % Column formatting, @{} suppresses leading/trailing space
      \toprule
      %\multicolumn{2}{c}{Item} \\
      \cmidrule(r){1-2} % Partial rule. (r) trims the line a little bit on the right; (l) & (lr) also possible
      Parameter    & Description & Value \\
      \midrule
      d$_c$ & low frequency edge band-defining diameter & 725 \micron \\
      d & waveguide diameter & 814~\micron \\
      z$_{\mathrm{bs}}$ & distance to backshort & 330~\micron \\
      g$_1$ & gap to waveguide interface & 25~\micron \\
      g$_2$ & gap to backshort & 25~\micron \\
      r$_1$ & choke width & 260~\micron \\
       & probe width & 228~\micron \\
       & probe height & 248~\micron \\
       & CPW width & 2~\micron \\
       & CPW gap & 4.5~\micron \\
            \bottomrule
   \end{tabular}
   \caption{Dimensions used in OMT-coupling simulations.}
   \label{tab:omt_dimensions}
\end{table}
   
The 280~GHz planar OMT has been scaled from existing designs optimized for the 150~GHz band \cite{mcmahon2009planar} and further refined with 3D electromagnetic simulations.
The model, presented in \figref{fig:omt_model} with dimensions listed in \tabref{tab:omt_dimensions}, consists of the low-frequency-edge defining waveguide of diameter d$_{c}$~=~725~\micron\, flared to diameter d~=~814~\micron\ and coupled to the three-stack silicon OMT-probe-coupled circuit. 
Insertion of the device membrane into the waveguide produces a waveguide gap of size g$_1$+g$_2$ (see \figref{fig:omt_model}b), which sources leakage radiation if not addressed.  
To mitigate, we include a quarter-wave transformer to help confine the waves to the waveguide.  
The surrounding four edges of the model are defined as radiation boundaries to capture and assess this leakage radiation.

The Nb probe dimensions are linearly scaled from the 150~GHz design.   
These lie on a dielectric stack that consists of a 0.45~\micron\ thick layer of SiO$_{\mathrm{x}}$ and 1~\micron\ thick layer of SiN$_{\mathrm{x}}$.
We model the relative permeability of the SiO$_{\mathrm{x}}$ and SiN$_{\mathrm{x}}$ as  $\epsilon_r$~=~4 and 7 respectively.  
The probes act as an impedance transformer between the $\sim$~600~$\Omega$ waveguide impedance (at 280~GHz) and the 117~$\Omega$ CPW transmission line, which is the output of the OMT.  
%Silicon underneath the CPW and outside of the choke radius has been removed in the model, since the silicon dielectric half space would alter the impedance.
  
The backshort distance z$_{\mathrm{bs}}$ has been chosen to maximize the band-averaged coupling.  
We find z$_{\mathrm{bs}}~=~\lambda_{\mathrm{g}}$/4.78, where $\lambda_{\mathrm{g}}$ is the guided wavelength at 280~GHz.  
The reason the optimum backshort distance differs from $\lambda_{\mathrm{g}}$/4 is due to the reactance of the Nb probes.  

In simulations from 240--330~GHz in 10~GHz steps, we launch TE11$_{\mathrm{x}}$, TE11$_{\mathrm{y}}$, and TM01 waveguide modes and determine the scattering parameters of the five waveports: 
the four OMT-CPW outputs and the input waveguide.  
As this is a lossless model, leakage radiation out of the waveguide choke is determined by the remaining power that is not captured in the 5 ports.    
\figref{fig:omt_model}c shows co-polar coupling, reflection, leakage radiation, and coupling to the higher order TM01 mode across the band.  
The design achieves 94\% band-averaged (252-315~GHz) co-polar coupling, 3\% reflection and 3\% leakage radiation.  
Coupling to the higher order TM01 mode, which is a source of cross-polarization and angular resolution degradation, is low ($<-40$~dB).  

%Cross-polar coupling arises from waveguide to probe mis-alignment and is limited to \comment{-XX~dB} with our alignment approach.
  
%%%%%%%%%%%%%%%%%%%%%%%%%%%%%%%%%%%%%%%%%%%%%%%%%%%%
\subsubsection{Co-Planar Waveguide (CPW) to Microstrip (MS) Transition}
\begin{figure*}[t]
\begin{overpic}[width=6.5in]{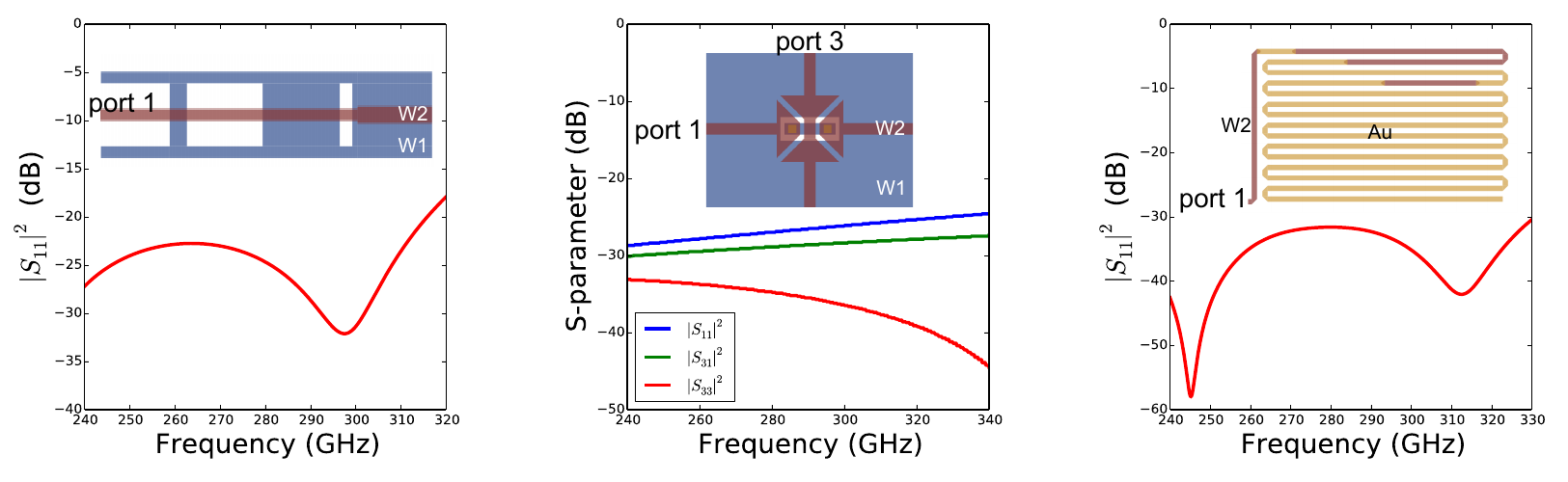}
\put(0,27){\normalsize \textcolor{black}{\bf (a)}}
\put(33,27){\normalsize \textcolor{black}{\bf (b)}}
\put(67,27){\normalsize \textcolor{black}{\bf (c)}}
\end{overpic}
\caption{ Microwave circuit element design and performance.
{\bf (a)} CPW to MS transition.
{\bf (b)} Microwave cross-under.
{\bf (c)} Load resistor coupling: Nb to Au microstrip transition.  (The continuous ground plane underneath these traces is not shown.)
W1 (blue) is a 200~nm thick Nb ground plane, W2 (red) is a 400~nm thick Nb layer and Au (yellow) is a 100~nm thick layer of gold.}
\label{fig:tline}
\end{figure*}

To confine the electromagnetic radiation, eliminating potential coupling of the fields to nearby structures, we choose to transmit power to the TES via microstrip (MS) transmission-line.  
Therefore, a Co-Planar Waveguide (CPW) to MS transmission line is required.  
Between the input 2/4.5~\micron\ width/gap CPW input and the 3~\micron\ wide, 350~nm thick SiN$_{\mathrm{x}}$ dielectric MS, 
we implement a 5-section transition presented in \figref{fig:tline}a.
The transition consists of alternate sections of 2/4.5~\micron\ CPW impedance and 2~\micron\ wide MS.  
The impedance of the CPW and MS are determined by use of a 2D electromagnetic simulator.  
These are inputs to a transmission line model then used to optimize the length of each section.
The full design is then validated in a 2D electromagnetic simulator.
Peak reflection across the band is $-23$~dB.

% Fig5: OMT model and results 

%%%%%%%%%%%%%%%%%%%%%%%%%%%%%%%%%%%%%%%%%%%%%%%%%%%%
\subsubsection{Microwave Cross-under}
We developed a microstrip cross-under to avoid the need for a third Nb layer.
Cutting the ground plane adds inductance to the transmission line.
We tune out this inductance by adding capacitive fins.  
Designs optimized over the band 120--270~GHz have been previously presented \cite{duff2016advanced}, and a similar design for a three-wiring layer cross-over has been independently developed \cite{posada2015fabrication}.
\figref{fig:tline}b presents the \spidertwo\ design, which has 3~\micron\ MS inputs and is optimized for the 280~GHz band.  
Reflection is below $-25$~dB for the signal lines, and the isolation between them is $<-28$~dB.

%%%%%%%%%%%%%%%%%%%%%%%%%%%%%%%%%%%%%%%%%%%%%%%%%%%%
\subsubsection{Termination Load Resistor}
\label{sec:loadR}
To deposit power from the Nb probe fins onto the bolometer, we transition from superconducting MS to lossy Au MS.  
An abrupt transition produces near 10\% reflection, and therefore we implement the transition presented in \figref{fig:tline}c that consists of alternating lengths of Nb MS and Au MS.
We assume $\sigma_{\mathrm{Au}}$~=~60.7~$\times$~10$^{6}$~S/m, determined from measurements at 4~K; a London penetration depth in Nb of 85~nm; and $\epsilon_{\mathrm{r}}$ (SiN$_{\mathrm{x}}$)~=~7.
Using a one-port simulation model, peak reflection is $-32$~dB.

%%%%%%%%%%%%%%%%%%%%%%%%%%%%%%%%%%%%%%%%%%%%%%%%%%%%
%%%%%%%%%%%%%%%%%%%%%%%%%%%%%%%%%%%%%%%%%%%%%%%%%%%%
\subsection{Transition-Edge Sensor (TES) Bolometer}
\label{loadR}

To sense power in each linear polarization, we use separate transition edge sensor (TES) bolometers \cite{irwin2005transition}.  
Photographs of the transition edge sensor bolometers, with labeled sub-components, are shown in \figref{fig:det_design}.  
The requirements of the bolometer are a saturation power \psat~=~3~pW and operation from 300~mK.
This saturation power is achieved with a bolometer transition temperature \tc~=~420~mK, and four thermally isolating silicon nitride legs, each of dimension 1000~\micron~$\times$~8~\micron~$\times$~1~\micron.  
The thermal conductance is G(\tc)~=~34~pW/K.  

Power from opposing Nb probes is deposited in the bolometer by use of two separate load resistor circuits as described in \secref{sec:loadR}.
As the sensing element, we utilize 0.25 squares of 1300~ppma, $\sim$~350~nm thick AlMn.  
This produces \tc~=~420~mK and a normal resistance \rn~=~10~m$\Omega$.    
TES implementations of AlMn have previously been shown for several SQUID multiplexing approaches and base operation temperatures \cite{deiker2004superconducting, schmidt2011almn, duff2016advanced}.  
As presented in Li et al.\cite{li2016almn}, the AlMn \tc\ is a function of composition and heating temperature.
To set \tc\ after film deposition, we bake the wafer at $\sim$~240~$^{\circ}$C, a temperature greater than the film will be exposed to in subsequent wafer processing.  

Like the \spiderone\ bolometers \cite{bk2015antenna}, the bolometer includes a higher \tc\ TES in series to extend the bolometer dynamic range for in-lab testing.
This feature is extremely enabling for a balloon experiment, where assessing any TES detector functionality immediately before flight is otherwise challenging. 
We implement a 0.25 square Al sensor, simply wired in series with the AlMn sensor using Nb traces.  
Given the leg geometry, which is defined for high sensitivity observations under in-flight loading, we require \tc~$\sim$~1.6~K to avoid saturation under 300~K loads.  
Thus the \tc\ from pure, bulk Al is too low, and we therefore dope the films with oxygen in order to raise the \tc.  
When performing science grade observations, the Al sensor is superconducting and as such is effectively absent from the bias circuit.

We control the electro-thermal response time to $\sim$~1~ms with symmetric PdAu volumes added to the bolometer island.
The bolometer has been designed based on a study of low thermal conductance prototype bolometers.

%%%%%%%%%%%%%%%%%%%%%%%%%%%%%%%%%%%%%%%%%%%%%%%%%%%%
%%%%%%%%%%%%%%%%%%%%%%%%%%%%%%%%%%%%%%%%%%%%%%%%%%%%
%%%%%%%%%%%%%%%%%%%%%%%%%%%%%%%%%%%%%%%%%%%%%%%%%%%%
\section{Balloon-optimized TES Bolometers}
\label{sec:bolo_study}

% Figure 6: dark bolometer performance
\begin{figure*}[t]
\begin{overpic}[height=2.5in]{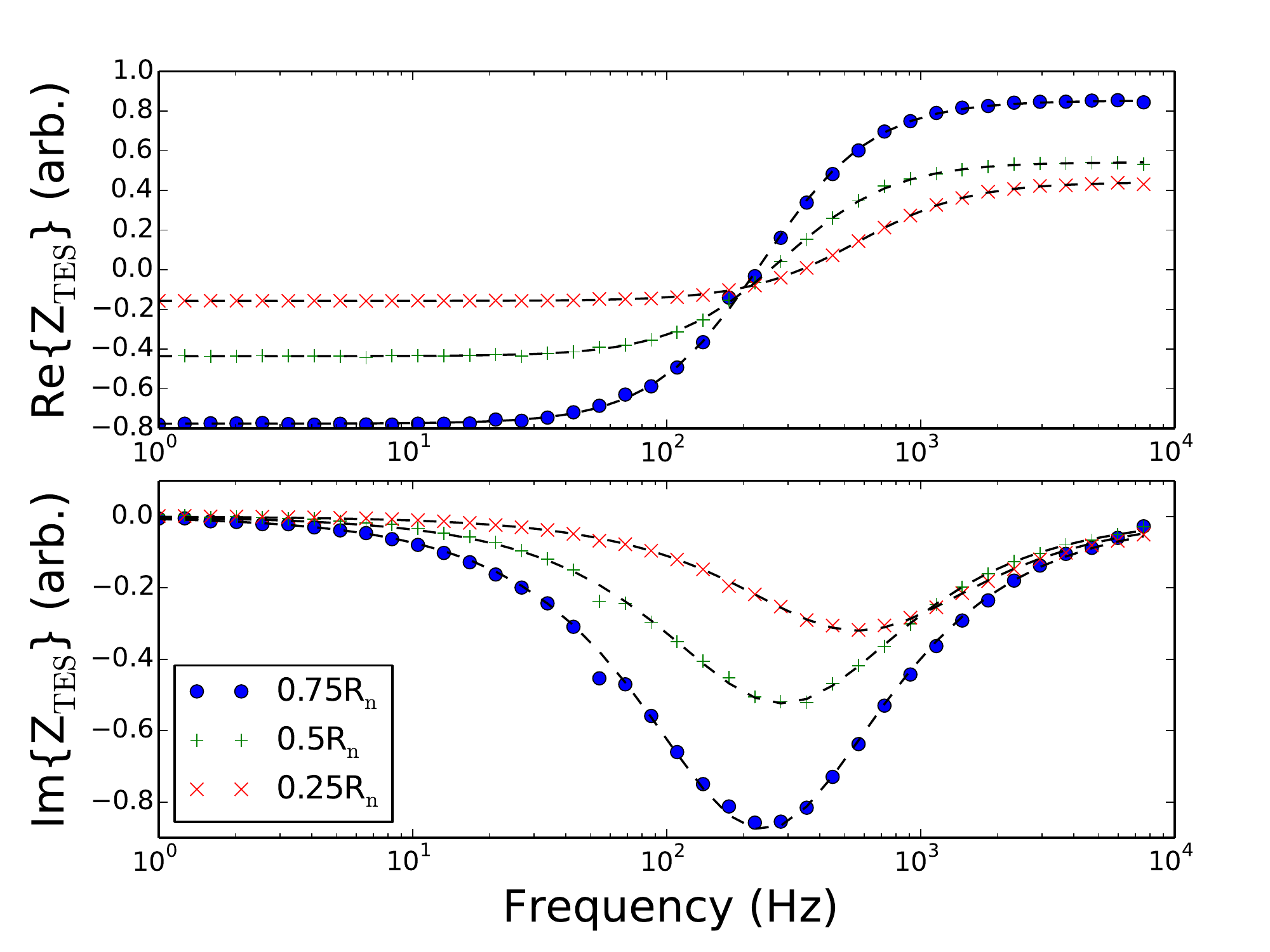}
\put(0,75){\normalsize \textcolor{black}{\bf (a)}}
\end{overpic}
\begin{overpic}[height=2.5in]{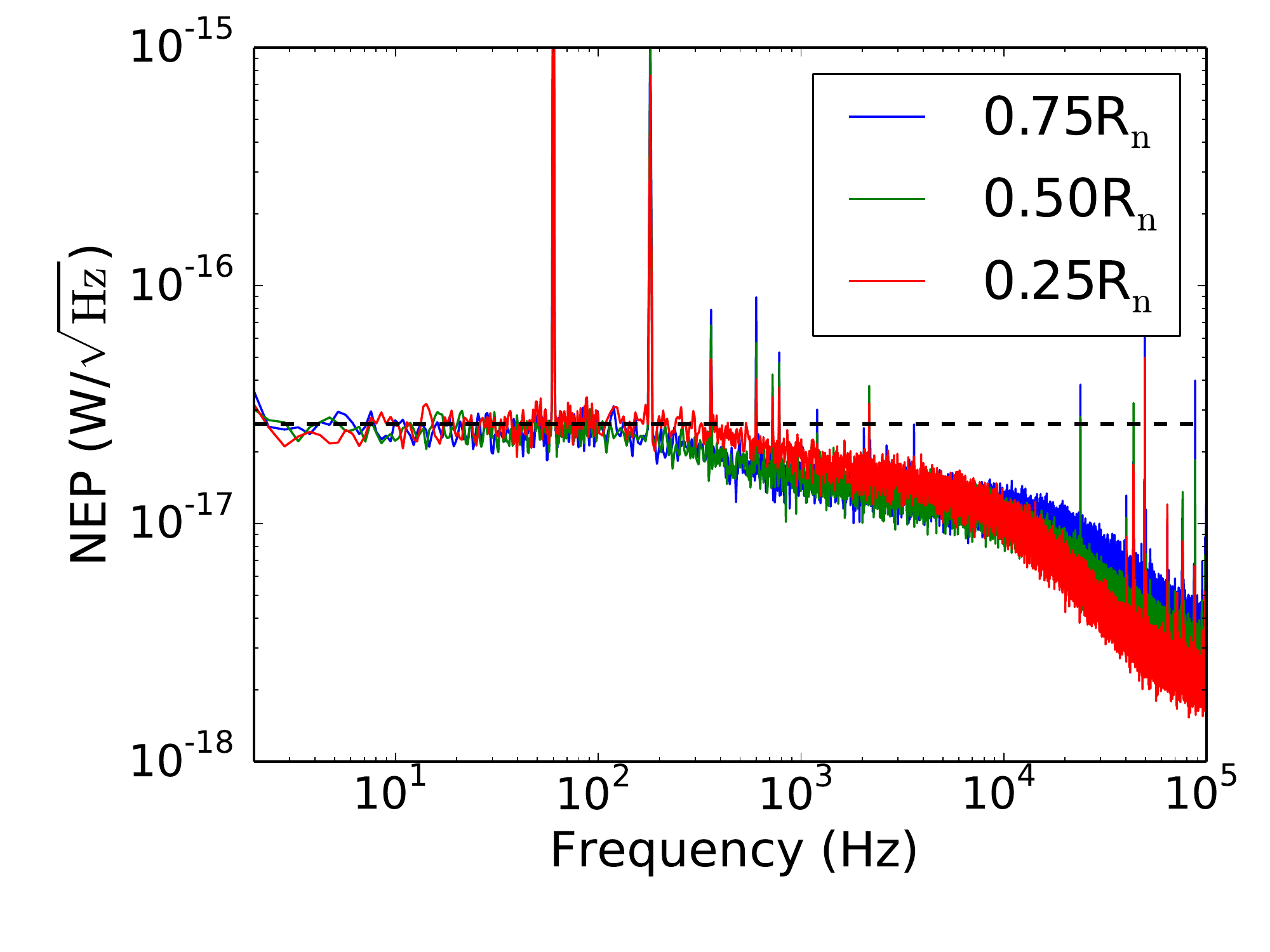}
\put(0,75){\normalsize \textcolor{black}{\bf (b)}}
\end{overpic}
\caption{
Prototype bolometer measurements at bias points 0.75\rn, 0.50\rn, and 0.25\rn.
{\bf (a)} The good fit of these complex impedance measurements to a one-pole response model demonstrates that the device is well-described by a simple TES bolometer model \cite{lindeman2004impedance,irwin2005transition}.
{\bf (b)} Electrical NEP measurements show that the white noise level is consistent with the thermal noise prediction \cite{boyle1959performance} (dashed black line).}
\label{fig:boloResults}
\end{figure*}   

To achieve high instantaneous sensitivity, taking advantage of the low background loading of a space-like environment, 
the bolometer saturation power \psat\ (or equivalently its thermal conductance G) must be substantially decreased as compared to a ground-based observing optimized bolometer. 
Using SiN$_x$ thermally isolating legs, \psat\ scales as $\sim$~\tc$^3$.  
Since \spider\ operates from a 300~mK base temperature, this necessarily dictates long thermally isolating bolometer legs.  
A suitable bolometer has been already been demonstrated \cite{bk2015antenna}.
In the Caltech/JPL design, the bolometer footprint is compact by implementing meandered, wider legs that support the transmission lines from the antenna.  
These lines are mechanically supported by thinner, straight legs.
Since the feedhorns create large areas on the detector wafer for circuit elements, we simply implement straight-leg bolometer isolation.

To aid the \spidertwo\ bolometer design, we fabricated bolometers with identical islands, but with different cross-sectional area to length (A/$\ell$) leg ratios.
The island design is similar to that presented in \figref{fig:det_design} but has no added PdAu, no series Al TES, and uses a full square of AlMn.
The most extreme A/$\ell$ ratio was A/$\ell$~= 24~\micron$^2$/1500~\micron\ (all four legs).  
Although these bolometers yielded the target \psat\ (with \tc~=~545~mK), the mechanical yield was less than 100\%.  
However, bolometers with A/$\ell$~= 24~\micron$^2$/1000~\micron\ did have 100\% yield. 
We therefore choose to pursue the shorter leg design, and we provide additional margin in the devices on the arrays by widening each leg to 8~\micron.
In order to achieve the \psat\ target, we compensate for the increased A/$\ell$ by lowering \tc.    

By measuring I-Vs as a function of bath temperature, we characterize the thermal conductance of the A/$\ell$~= 24~\micron$^2$/1000~\micron\ bolometer.  
Fitting to the functional form \psat~=~$K$(\tc$^n$-\tb$^n$) \cite{irwin2005transition}, we find $K$~=~70~pW/K$^n$, $n$~=~3.5 and \tc~=~545~mK.  
We then assess the dynamics and sensitivity by measuring complex impedance and noise at multiple voltage bias positions.  Three are shown in the figure: 0.75\rn, 0.5\rn, 0.25\rn.   
For these measurements, we elevate the bath temperature to 480~mK, so as to achieve \psat~=~3~pW.  
\figref{fig:boloResults} shows the results.  
For the impedance measurements, the bias circuit has been removed and the data normalized by use of the superconducting and normal state impedance measurements, following Lindeman et al \cite{lindeman2007complex}.
The real and imaginary parts of the impedance are displayed together with one-pole fits.  
The good agreement between the data and model for every bias position demonstrates that the bolometer is isothermal and well-described by a simple TES bolometer model \cite{lindeman2004impedance, irwin2005transition}.  
From these measurements we infer the heat capacity C~=~0.8~$\pm$~0.1~pJ/K, the loopgain $\mathcal{L} > 20$ for a large fraction of the transition, and the logarithmic sensitivity of the resistance to 
current $\beta~<~1$, throughout the transition. 

\figref{fig:boloResults}b shows the electrical noise equivalent power (NEP).
To convert from the measured current noise to electrical NEP, the detector voltage bias has been assumed for the device 
responsivity\footnote{The frequency-dependence of the responsivity has been neglected, and therefore NEP values above $\sim$~300~Hz are misleading.  They are presented to show the frequency roll-off only.}. 
The measured white noise NEP~=~2.6~$\times~10^{-17}$~W$^2$/Hz matches the thermal noise prediction of Boyle and Rogers \cite{boyle1959performance}, calculated using the 
thermal conductance parameters $K$, $n$, and \tc\, which have been determined from the I-V curves.  
No excess noise is evident at high modulation frequencies, further evidence to support the simple TES model interpretation.  
%The data do show increased noise at low frequencies.
%However we note that there is no polynomial subtraction of the raw data, and the measurement has not been optimized for low frequency noise investigations.  
%We have previously demonstrated 10~mHz 1/$f$ knees in pair-differenced, space-optimized bolometers \cite{niemack2012optimizing}.

%%%%%%%%%%%%%%%%%%%%%%%%%%%%%%%%%%%%%%%%%%%%%%%%%%%%
%%%%%%%%%%%%%%%%%%%%%%%%%%%%%%%%%%%%%%%%%%%%%%%%%%%%
%%%%%%%%%%%%%%%%%%%%%%%%%%%%%%%%%%%%%%%%%%%%%%%%%%%%
\section{Conclusions}
\label{sec:conclusions}
We have described the design of 280~GHz bolometric arrays for \spidertwo.  
Electromagnetic simulations show high-performance, polarimetric optical coupling.  
The fabrication of both the designed feedhorns and detectors has been demonstrated on single pixels.  
Measurements of these devices prove several aspects of the design, including low thermal conductance bolometers and the feedhorn corrugation profile.  
Three arrays of this design instrument \spidertwo.  
These polarized dust monitoring channels aid in the search for inflationary gravity waves, hidden in the polarization of the \cmb.

%%%%%%%%%%%%%%%%%%%%%%%%%%%%%%%%%%%%%%%%%%%%%%%%%%%%
%%%%%%%%%%%%%%%%%%%%%%%%%%%%%%%%%%%%%%%%%%%%%%%%%%%%
%%%%%%%%%%%%%%%%%%%%%%%%%%%%%%%%%%%%%%%%%%%%%%%%%%%%
\acknowledgments % equivalent to \section*{ACKNOWLEDGMENTS}       
The \spider\ project is supported by NASA under APRA
grant NNX12AE95G, issued through the Science Mission Directorate, by the
NSF through the award PLR-1043515, and by the David and Lucile Packard
Foundation.  Logistical support for the Antarctic deployment and
operations was provided by the NSF through the U.S. Antarctic Program.  
We performed our 3D electromagnetic simulations on the Baker-Jarvis Cluster 
and supporting infrastructure in the Communications Technology Laboratory at NIST, Boulder.

% References
%\bibliography{spie} % bibliography data in report.bib
%\bibliographystyle{spiebib} % makes bibtex use spiebib.bst

\end{document}